%% file: ojcoms.tex
\documentclass{IEEEoj}

\usepackage{silence}
\usepackage{cite}

\usepackage{amsmath,amssymb,amsthm, amsfonts, fixmath}
\usepackage{pifont}% http://ctan.org/pkg/pifont
\usepackage{xcolor}
\usepackage{colortbl} % Required for \columncolor

\let\labelindent\relax
\usepackage[inline]{enumitem}
\usepackage{amsmath}
\usepackage{algorithm}
\usepackage{algpseudocode}
\usepackage{graphicx}
\usepackage{float}
\usepackage[utf8]{inputenc}
\usepackage{siunitx}
\DeclareSIUnit{\dBm}{dBm}
\DeclareSIUnit{\dBW}{dBW}

\usepackage{graphicx}
\usepackage[labelformat=simple]{subcaption}

\usepackage{epstopdf}
\usepackage{lipsum}

\usepackage{cuted}
\usepackage{multirow,booktabs}
\usepackage{diagbox}

\usepackage{optidef}

\usepackage[hidelinks]{hyperref}

\usepackage[shortcuts,acronym,automake]{glossaries}
\input{glossary}

\usepackage{tcolorbox}
\tcbuselibrary{many}
\newcommand{\subscript}[1]{_{\mathrm{#1}}}

\newcommand{\revise}[1]{#1}

\newcommand{\revboxtwo}[1]{#1}
\newcommand{\revthree}[1]{{\color{\highlightcolor}#1}}
\renewcommand{\revthree}[1]{#1}

\usepackage[textsize=tiny,colorinlistoftodos]{todonotes}
\makeatletter
\define@key{todonotes}{bh}[]{% Comment by 
	\setkeys{todonotes}{author=\textbf{Bin}, color=cyan!30}}%
\define@key{todonotes}{yy}[]{% Comment by 
	\setkeys{todonotes}{author=\textbf{Ye}, color=green!30}}%
\makeatother

\newif\ifapp
\appfalse
\apptrue

\def\BibTeX{{\rm B\kern-.05em{\sc i\kern-.025em b}\kern-.08em
    T\kern-.1667em\lower.7ex\hbox{E}\kern-.125emX}}
\AtBeginDocument{\definecolor{ojcolor}{cmyk}{0.93,0.59,0.15,0.02}}
\def\OJlogo{\vspace{-4pt}\hskip-4pt\includegraphics[height=18pt]{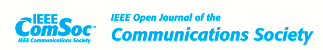}}
\begin{document}
\receiveddate{XX Month, XXXX}
\reviseddate{XX Month, XXXX}
\accepteddate{XX Month, XXXX}
\publisheddate{XX Month, XXXX}
\currentdate{19 January, 2026}
\doiinfo{OJCOMS.2026.XXXXXX}

\title{VENENA: A Deceptive Visual Encryption Framework for Wireless Semantic Secrecy}

\author{BIN HAN\IEEEauthorrefmark{1} \IEEEmembership{(Senior~Member, IEEE)}, YE YUAN \IEEEauthorrefmark{2}\IEEEmembership{(Member, IEEE)},AND HANS D. Schotten\IEEEauthorrefmark{1,3}
\IEEEmembership{(Member, IEEE)}}
\affil{RPTU University Kaiserslautern-Landau, 67663 Kaiserslautern, Germany}
\affil{Tongji University, 201804 Shanghai, China}
\affil{DFKI GmbH, 67663 Kaiserslautern, Germany}
\corresp{CORRESPONDING AUTHOR: Bin Han (e-mail: bin.han@rptu.de).}
\authornote{This work is supported in part by the European Commission via the Horizon Europe project Hexa-X-II (101095759), and in part by the Federal Ministry of Research, Technology and Space of Germany via the projects 6G-ANNA (16KISK105) and Open6GHub+ (16KIS2406).}
\markboth{Preparation of Papers for IEEE OPEN JOURNAL OF THE COMMUNICATIONS SOCIETY}{Han \textit{et al.}}

\begin{abstract}
% Eavesdropping has been a long-standing threat to the security and privacy of wireless communications, since it is difficult to detect and costly to prevent. As networks evolve towards \ac{6g} and semantic communication becomes increasingly central to next-generation wireless systems, securing semantic information transmission emerges as a critical challenge. While classical \ac{pls} focuses on passive security, the recently proposed concept of \ac{pld} offers a semantic encryption measure to actively deceive eavesdroppers. Yet the existing studies of \ac{pld} have been dominantly information-theoretical and link-level oriented, lacking considerations of system-level design and practical implementation.

% In this work we propose a novel \ac{ai}-enabled framework called \ac{venena}, which combines the techniques of \ac{pld}, visual encryption, and image poisoning, into a comprehensive mechanism for deceptive secure semantic transmission in future wireless networks. By leveraging advanced vision transformers and semantic codecs, \ac{venena} demonstrates how semantic security can be enhanced through the synergy of physical layer techniques and artificial intelligence, paving the way for secure semantic communication in \ac{6g} networks.

{Eavesdropping has been a long-standing threat to the security and privacy of wireless communications, since it is difficult to detect and costly to prevent. As networks evolve towards \ac{6g} and semantic communication becomes increasingly central to next-generation wireless systems, securing semantic information transmission emerges as a critical challenge. While classical \ac{pls} focuses on passive security, the recently proposed concept of \ac{pld} offers a semantic encryption measure to actively deceive eavesdroppers. Yet the existing studies of \ac{pld} have been dominantly information-theoretical and link-level oriented, lacking considerations of system-level design and practical implementation.

In this work we propose \ac{venena}, an artificial intelligence-enabled framework for secure image-based communication. \ac{venena} protects confidential messages by encoding them visually while actively deceiving eavesdroppers: legitimate receivers use \ac{ai}-based classifiers to extract true message semantics, while interceptors perceive only falsified content. The framework transmits two superimposed image components with different power levels--a high-power decoy image and a low-power correction mask--ensuring only authorized receivers with favorable channel conditions can reconstruct the true message. Experimental validation demonstrates over 93\% accuracy for legitimate users while limiting eavesdropper success to 52\% even when system design is fully known, validating \ac{venena}'s active defense capability for \ac{6g} semantic communication.}
\end{abstract}


\maketitle

\glsresetall

\section{INTRODUCTION}\label{sec:introduction}
% \todo[bh,inline]{Revise the abstract, intro, and conclusion to emphasize that \ac{pld} is a semantic security solution, so is VENENA.}
While the rapid development and evolution of wireless technologies have contributed to the latest \ac{5g} mobile systems, and leading us towards the future \ac{6g} networks, the imperative for robust security and privacy measures has also become increasingly pronouced. As outlined in \cite{JHH+2021road}, the rapid pace of technological advancements has complicated the cybersecurity landscape, necessitating transformative security strategies to counteract both existing and emerging threats effectively.

One of the major concerns in wireless security and privacy is that the classical cryptographic algorithms are no more guaranteed to be secure in the future. On the one hand, the recent advent of quantum computing poses a significant threat to such methods of computational security. 
% On the other hand, the extensive softwarization, virtualization, and cloudification of network functions in \ac{5g} and future \ac{6g} networks, often in a multi-tenant environment, introduce new vulnerabilities and attack surfaces, increasing the risk of data leakage, which further degrades the security provided by classical cryptographic methods, which fundamentally rely on the data confidentiality of authentication information. 
{On the other hand, the extensive softwarization, virtualization, and 
cloudification of network functions in \ac{5g} and future \ac{6g} networks introduce new vulnerabilities in multi-tenant environments. These increase the risk of data leakage, further degrading the security provided by classical cryptographic 
methods that fundamentally rely on confidential authentication information.}
In this context, the concept of \ac{pls} has emerged as a promising alternative to traditional cryptographic methods. Leveraging the physical properties of the wireless channel to provide secure communication, \ac{pls} offers a new paradigm from preventing eavesdropping attacks and ensuring data confidentiality.

However, despite %the enhancements it brings in passive security, conventional \ac{pls} technologies play no better than classical cryptgraphic security measures regarding the lack of capability to actively defend against eavesdroppers, i.e. to decourage, detect, or damage the eavesdroppers. 
{enhancing passive security, conventional \ac{pls} technologies cannot actively defend against eavesdroppers--they lack the capability to discourage, detect, or disrupt adversaries. In this regard, \ac{pls} performs no better than classical cryptographic security measures.}
To bridge this gap, 
% we have proposed in \cite{CHZ+2025physical} a novel framework of \ac{pld}, which uses \ac{pls} to realize a deceptive defense against eavesdroppers. 
{the concept of \ac{pld} was first proposed in~\cite{HZS+2023non}, which introduced deceptive signaling via non-orthogonal multiplexing to mislead eavesdroppers under finite blocklength constraints. This idea was later extended in~\cite{CHZ+2025physical} to derive secrecy capacity bounds and optimal power-allocation strategies, and further applied to \ac{ofdm} systems in~\cite{CHZ+2025physical2}.}
By deceiving eavesdroppers with falsified information, while simultaneously transmitting the true information to legitimate receivers, \ac{pld} provides an active defense mechanism to enhance the security of wireless communications. 

{Especially, the practical value of \ac{pld} emerges in scenarios where adversaries can suffer from inaccurate information. For example, in military tactical communications, forward units can transmit target coordinates as deceptively-encoded messages: intercepting adversaries decode falsified coordinates that misdirect their resources, while legitimate receivers reconstruct true coordinates. In confidential business communications, as another instance, corporate executives can transmit merger target identifiers or product launch schedules in the same way: malicious insiders or competitors with leaked protocols intercept falsified information that triggers counterproductive actions, while authorized partners receive true intelligence. Such active deception proves more damaging to adversaries than signal denial, as they commit resources based on false intelligence rather than merely lacking information.}
We have later proven that \ac{pld} is a semantic encryption solution for secure transmission, which relies on the secrecy of the semantic knowledge instead of that of the messages~\cite{HZS+2025semantic,CHZ+2025physical3}. \revthree{Compared to contemporary works in adversarial semantic communication that focus on the \emph{attack-oriented} generation of semantic perturbations--such as SemAdv~\cite{NLZ+2023physical}, SemBLK~\cite{LLN+2023boosting}, and SemPerGe~\cite{AMA+2025semperge}--or on protecting semantic communication systems against such perturbations--such as \emph{SemProtector}~\cite{LNC+2023semprotector}--\ac{pld} targets a fundamentally different objective. Specifically, rather than suppressing semantic distortion, \ac{pld} \emph{discriminatively induces} semantic distortions at illegitimate eavesdroppers while preserving semantically correct perception at the legitimate receiver.}

As the core part of the \ac{pld} framework, the deceptive ciphering algorithm is required to fulfill two specific requirements. First, the codebooks of ciphertext and plaintext have to share a same codewrd set, i.e. every possible ciphertext codeword must also be a valid plaintext codeword. Second, every individual codeword must contain meaningful semantic information. While the preliminary works on \ac{pld} \revthree{\cite{HZS+2023non,CHZ+2025physical,CHZ+2025physical2,HZS+2025semantic,CHZ+2025physical3} have provided in-depth information-theoretic and semantic analyses--on perspectives of deception rate, leakage-failure rate, and semantic distortion--as the cornerstone, they have}
been only focusing on the link-level design, little has been discussed about the application scenario or the system-level design, which can be constrained by the above requirements. 
% Furthermore, existing \ac{pld} studies lack the essential semantic encoding mechanisms needed to generate the meaningful semantic codewords required by the framework, creating
{essential semantic encoding mechanisms to generate meaningful semantic codewords. This creates} a critical implementation gap between theoretical \ac{pld} concepts and practical semantic communication systems.

\revise{Visual encryption provides the essential semantic encoding capability that bridges this gap between theoretical \ac{pld} requirements and practical implementation. By mapping confidential messages onto semantically meaningful visual content, visual encryption enables the creation of dual-purpose codewords that satisfy both \ac{pld}'s semantic richness requirement and the deception objectives of active security. However, traditional visual encryption approaches remain fundamentally passive, lacking the active deception capabilities needed to counter sophisticated eavesdroppers who may possess full system knowledge.}

As a proof of concept for \ac{pld}, in this work we propose a deceptive visual encryption framework called \ac{venena}, which represents the first system-level implementation of \ac{pld} theory by combining \ac{pld} with image poisoning to realize secure and deceptive transmission of secrecy messages over wireless wiretap channels. \ac{venena} advances the field by:
% \begin{enumerate}
%     \item providing the first practical framework that translates theoretical \ac{pld} concepts into a working system for semantic communication; 
%     \item introducing \ac{ai}-enabled image poisoning as the key enabler for generating semantically meaningful deceptive content; and 
%     \item demonstrating how the synergy of physical layer techniques and artificial intelligence can enhance semantic security in next-generation wireless networks.    
% \end{enumerate}
{
\begin{enumerate}
    \item \textbf{System-level instantiation}: While prior \ac{pld} work \cite{CHZ+2025physical,HZS+2025semantic} established information-theoretic foundations deriving secrecy capacity bounds under the assumption that de- ceptive ciphering algorithms exist, no concrete implementation of such algorithms or system architecture was provided. \ac{venena} provides the first practical realization of \ac{pld} by addressing: 
    \begin{enumerate*}[label=(\roman*)]
        \item semantic encoding via visual classification networks, 
        \item deceptive content generation through gradient matching attacks, and 
        \item power-domain multiplexing for dual-message transmission,
    \end{enumerate*}
    transitioning \ac{pld} from theoretical abstraction to engineered system;
    \item \textbf{\Ac{ai}-enabled deceptive content generation}: We introduce \ac{gma}-based image poisoning as the realization of \ac{pld}'s deceptive ciphering requirement, enabling adversarial perturbations that maintain visual similarity while achieving semantic transformation. This addresses the critical gap between 
   \ac{pld}'s requirement for ``semantically meaningful codewords'' and practical generation mechanisms;
   \item \textbf{Empirical validation against informed adversaries}: We demonstrate quantifiable security gains ($>40\%$ reduction in eavesdropper accuracy) even when adversaries possess full system knowledge, validating \ac{pld}'s active defense capability against insider threats that defeat traditional visual encryption.
\end{enumerate}

An overview of \ac{venena}'s features compared to conventional \ac{pls} and existing \ac{pld} works is provided in Table~\ref{tab:overveiw_venena}.
}
\begin{table*}[!htbp]
    \centering
    % \captionsetup{font={color=\highlightcolor}}
    \caption{Overview of \ac{venena} features compared to conventional \ac{pls} and existing \ac{pld} works.}\label{tab:overveiw_venena}
    {
    \setlength{\extrarowheight}{5pt}% (optional, requires the array package) 
    \begin{tabular}{p{2.5cm} p{4.5cm} p{4.5cm} p{4.5cm}}
        \toprule[2px]
        &   \textbf{Conventional \ac{pls}}  &   \textbf{Existing \ac{pld}}  &   \textbf{\ac{venena}} \\
        \midrule[1px]
        \textbf{Active defence} &   No  &   Yes  &   Yes\\
        % &&&\\
        % \hline
        \textbf{Basis} &   Randomness of the physical channel  &   Statistical channel superiority of the legitimate receiver  over eavesdropper    &    Statistical channel superiority of the legitimate receiver over eavesdropper\\
        % &&&\\
        % \hline
        \textbf{Semantic encryption} &   Irrelevant  &   \revthree{Necessary}, unspecified &    \ac{ai}-driven visual encryption \\
        % &&&\\
        % \hline
        \textbf{System-level demonstration} &   Reported   &   Unavailable so far  &   Yes (first practical \ac{pld} framework) \\
        \bottomrule[2px]
    \end{tabular}
    }
\end{table*}

The remainder of this article is organized as follows. In Sec.~\ref{sec:related} we provide a brief tutorial to the related works, outlining the state-of-the-art in the fields of \ac{pls}, cyber deception, \ac{pld}, visual encryption, and image poisoning. In Sec.~\ref{sec:framework} we introduce the proposed \ac{venena} framework, detailing its principles and design. As the technical enabler, the image poisoning algorithm is presented in Sec.~\ref{sec:poisoning}, with a step-by-step description of the poisoning process. Subsequently, in Sec.~\ref{sec:simulation} we present the numerical verification of the \ac{venena} framework, demonstrating its effectiveness in secure and deceptive wireless image transmission. 
% Finally, in Sec.~\ref{sec:conclusion} we conclude the article and provide an outlook on future research directions.
\revise{Some complementary discussions regarding the efficiency aspects of our proposed solution are made then in Sec.~\ref{sec:discussion}. In the end, with Sec.~\ref{sec:conclusion} we conclude the article and give a glance into future research possibilities.}

\section{RELATED WORKS}\label{sec:related}
\subsection{PHYSICAL LAYER SECURITY}\label{subsec:pls}
% The concept of \ac{pls} originates from the seminal work of \emph{Wyner}~\cite{Wyner1975wiretap}, which generalizes \emph{Shannon}'s concept of perfect secrecy~\cite{Shannon1949communication} into a measurable strong secrecy over wiretap channels. Since then, the secrecy performance of communication systems has been studied over different variants of wiretap channels, including binary symmetric channels~\cite{CH1977note}, degraded \ac{awgn} channels~\cite{CH1978gaussian}, fading channels~\cite{GLE2008secrecy}, multi-antenna channels~\cite{ZGA2010secure}, broadcast channels~\cite{CK1978broadcast}, multi-access channels~\cite{TY2008gaussian}, interference channels~\cite{LSP+2009capacity}, relay channels~\cite{Oohama2007capacity}, etc. \cite{MFH+2014principles,WBZ+2019survey}, to enhance the secrecy performance of communication systems.

Research on \ac{pls} can be traced back to \emph{Wyner}'s foundational research~\cite{Wyner1975wiretap}, which expanded \emph{Shannon}'s perfect secrecy theorem~\cite{Shannon1949communication} by introducing quantifiable strong secrecy metrics for wiretap channel scenarios. 
This breakthrough led to extensive studies examining secrecy characteristics across diverse wiretap channel configurations%
% , including symmetric binary channels, channels with degradation, systems with fading effects, multi-antenna setups, broadcast environments, multiple-access scenarios, channels with interference, and relay-based architectures. 
{: symmetric and degraded channels, fading systems, multi-antenna and relay architectures, and broadcast/multiple-access scenarios.}
Building upon this theoretical foundation, comprehensive surveys have systematically classified \ac{pls} techniques across different wireless technologies. The work in \cite{HFA+2024classifications} provides a complete taxonomy of physical layer security approaches for confidentiality, categorizing techniques from information-theoretic perspectives to practical implementations in modern wireless systems. Similarly, \cite{WBZ+2019survey} offers an extensive analysis of optimization approaches specifically targeting wireless physical layer security, encompassing resource allocation strategies, precoding techniques, and cooperative protocols. 
The theoretical understanding gained from these investigations has enabled
% the development of various security-enhancing techniques, encompassing resource allocation in radio networks, precoding and beamforming strategies, selective approaches for antennas and nodes, cooperative protocols, and specialized channel coding methods, all aimed at strengthening the security capabilities of communication systems.
{various security-enhancing techniques: resource allocation in radio networks, precoding and beamforming strategies, antenna and node selection, cooperative protocols, and specialized channel coding. These techniques collectively strengthen communication system security.}
However, despite these advances, conventional \ac{pls} approaches remain fundamentally passive, lacking the capability to actively counter sophisticated eavesdroppers.

\subsection{CYBER DECEPTION}\label{subsec:cyber_deception}
% In the field of information security, the principles of deception were firstly introduced and well demonstrated by the infamous practices of social engineering by \emph{Mitnick}~\cite{MS2003art}. Later, this concept was transferred by \emph{Cheswick}~\cite{Cheswick1992evening} and \emph{Stoll}~\cite{Stoll2005cuckoo} into defensive applications, which were originally called \emph{honeypots} and thereafter generalized to a broader spectrum of \emph{deception technologies}~\cite{JFT+2024comprehensive}. The core principle of deception technologies is misleading and distracting potential attackers with fake targets, e.g. fabricated data with similar features like the confidential data, and therewith protecting genuine information. These technologies can also entice attackers into revealing themselves, offering a proactive approach to maintaining security integrity. Over the past three decades, deception technologies have been well developed and widely adopted in information systems, across the four layers of network, system, application, and data. Various solutions have been proposed to mitigate, prevent, or to detect cyber attacks. For a comprehensive review on the state-of-the-art of deception technologies, readers are referred to~\cite{JFT+2024comprehensive,HKB2018deception,PCZ2019game}. 

The application of deception in cybersecurity was first popularized through \emph{Mitnick}'s notable social engineering exploits~\cite{MS2003art}. This concept eventually evolved into defensive mechanisms, initially emerging as honeypot systems before expanding into a comprehensive suite of deception-based security solutions. \revise{The field has matured significantly over the past decades, with a recent comprehensive survey~\cite{JFT+2024comprehensive} providing detailed analysis of deception techniques specifically designed for honeypots and honeynets. The survey reveals that modern cyber deception technologies have evolved from simple decoy systems to sophisticated frameworks capable of adaptive behavior and real-time response to attackers.} These technologies operate on a fundamental strategy of deploying decoy targets, such as synthetic data designed to mimic confidential information, to misdirect and confuse potential attackers while safeguarding authentic assets. Additionally, these deceptive methods serve as effective tools for exposing adversaries, enabling a proactive security stance. Throughout the last thirty years, deception-based security has matured significantly, with implementations spanning across all major layers of information systems: network, system, application, and data. Researchers have developed numerous approaches to address cyber threats through detection, prevention, and mitigation strategies. \revise{The evolution from passive honeypots to active deception frameworks demonstrates the critical shift from merely detecting attacks to actively misleading and disrupting adversaries.} For an in-depth exploration of current developments in deception technologies, interested readers should consult~\cite{HKB2018deception}.

\subsection{SEMANTIC TRANSMISSION SECURITY}\label{subsec:semantic_transmission_security}
The emerging concept of semantic communication has started reshaping the understanding and designing of communication systems in various aspects. In the context of security, semantic communication can be leveraged by different means, such like semantic data encryption, covert communication, and quantum-empowered semantic security. A comprehensive survey is provided by~\cite{GWZ+2024survey}. 

\subsection{PHYSICAL LAYER DECEPTION}\label{subsec:pld}
Deception technologies at the wireless physical layer remain an emerging field of study. 
% While our earlier research~\cite{CHZ+2025physical}, as discussed in Sec.~\ref{sec:introduction}, explores this domain, 
{The original concept of \ac{pld} was first proposed in~\cite{HZS+2023non}, which introduced a non-orthogonal multiplexing-based approach to embed deceptive information directly into the physical layer, thereby misleading eavesdroppers under finite blocklength constraints. Subsequent work in~\cite{CHZ+2025physical} generalized this framework, establishing its theoretical foundations through secrecy-capacity analysis, while~\cite{CHZ+2025physical2}further demonstrated its applicability to \ac{ofdm} systems, bridging the gap between theoretical modeling and practical waveform design. More recently, \cite{HZS+2025semantic} established \ac{pld} formalized \ac{pld} as a semantic encryption mechanism, highlighting its potential for secure semantic communication in next-generation wireless networks.
In a closely related but independent line of research on waveform-level deception, similar ideas were explored in~ \cite{HFW+2023proactive}, where researchers exploit \ac{mimo} systems' spatial diversity to generate a secondary deceptive signal stream alongside the legitimate one. However, these studies primarily focus on spatial deception rather than semantic deception, and therefore lack the systematic framework necessary for practical deployment in semantic communication scenarios.} 

These {efforts} represent early steps {toward} applying deception principles at the physical layer, {but they do not address the semantic encoding requirements essential for meaningful and controllable deception in future 6G systems.}

{Despite these advances, existing \ac{pld} research remains predominantly information-theoretic, concentrating on link-level secrecy analysis under the assumption that deceptive ciphering algorithms exist with desired properties. No prior work has demonstrated a concrete architecture or implementation capable of generating semantically meaningful falsified content. The \ac{venena} framework fills this critical gap by providing the first system-level realization of \ac{pld} through AI-enabled image poisoning, power-domain multiplexing, and empirical performance validation. In contrast to the abstract and asymptotic analyses in previous \ac{pld} studies, \ac{venena} delivers a complete end-to-end system architecture, quantifiable performance results, and experimental verification under realistic channel conditions.}

\subsection{VISUAL ENCRYPTION}\label{subsec:sota_visual_encryption}
% \todo[bh,inline]{\textbf{1. Are you really sure that we are talking about the same concept of visual encryption here? I used the term for "embedding information into images like watermarks", but what is described here seems like the encryption of images\dots}}
% \todo[bh,inline]{2. Since we can have only up to 15 references, please keep only the necessary citations (and comment the removed ones out from the reference list).}
Visual encryption is a technique that aims to protect the confidentiality and privacy of visual information by embedding it into images in a visually unrecognizable form. The goal is to ensure that the encrypted image appears similar to the original image while concealing the embedded information. Various approaches have been proposed in the literature to achieve visual encryption through information embedding.

Digital watermarking is one of the most well-known techniques for embedding information into images \cite{podilchuk2001digital}. 
However, the amount of information that can be embedded into an image using digital watermarking is limited, and increasing the embedding capacity often comes at the cost of reduced robustness or visual quality.

{Optical transform-based approaches have also been explored to achieve visual encryption through mathematical transformations. For example, quaternion Fresnel transform combined with chaotic systems can process multiple images holistically and transform recognizable visual content into noise-like patterns suitable for watermarking and secure transmission~\cite{yu2018four}. While such methods provide strong obfuscation through mathematical operations, they fundamentally rely on key secrecy--when adversaries obtain system parameters through protocol leakage or insider access, the encrypted content appears merely as random noise without providing active misdirection.}

Recent advancements in visual encryption have leveraged deep learning techniques to embed information into images. Deep learning-based approaches utilize the power of convolutional neural networks to learn complex embedding and extraction mappings, enabling more robust and adaptive visual encryption schemes \cite{geiping2020witches}. \revise{However, these systems remained primarily defensive and lacked deceptive capabilities needed in adversarial environments. The critical breakthrough came with the integration of adversarial machine learning techniques, particularly image poisoning, which enables dual functionality: securely transmitting hidden information while actively misleading potential adversaries through carefully crafted perturbations.}
Image poisoning with embedded information, it becomes possible to transmit secret messages or watermarks securely while also deceiving potential adversaries \cite{zheng2022poisoning}. \revise{Unlike purely defensive approaches, poisoning techniques enable dual functionality: securely transmitting hidden information while actively misleading potential adversaries through carefully crafted perturbations. This advancement overcame the passive nature of previous methods, creating systems capable of both concealment and misdirection.}

\section{THE VENENA FRAMEWORK}\label{sec:framework}
We consider an wiretap channel model with a sender (\emph{Alice}), a legitimate receiver (\emph{Bob}), and an eavesdropper (\emph{Eve}). \emph{Alice} wants to transmit a short confidential message (e.g., a number) to \emph{Bob}, while preventing \emph{Eve} from correctly recovering the message. \revthree{We consider \emph{Eve} as able to continuously monitor the wireless channel and intercept all transmissions from \emph{Alice} to \emph{Bob}. It may have or not have full knowledge about the details of the system design. Both the legitimate and eavesdropping channels are lossy, yet with sufficiently high capacity to transmit an image at required data rate.}

\begin{figure*}
    \centering
    \begin{subfigure}[t]{\linewidth}
        \centering
        \includegraphics[scale=.55]{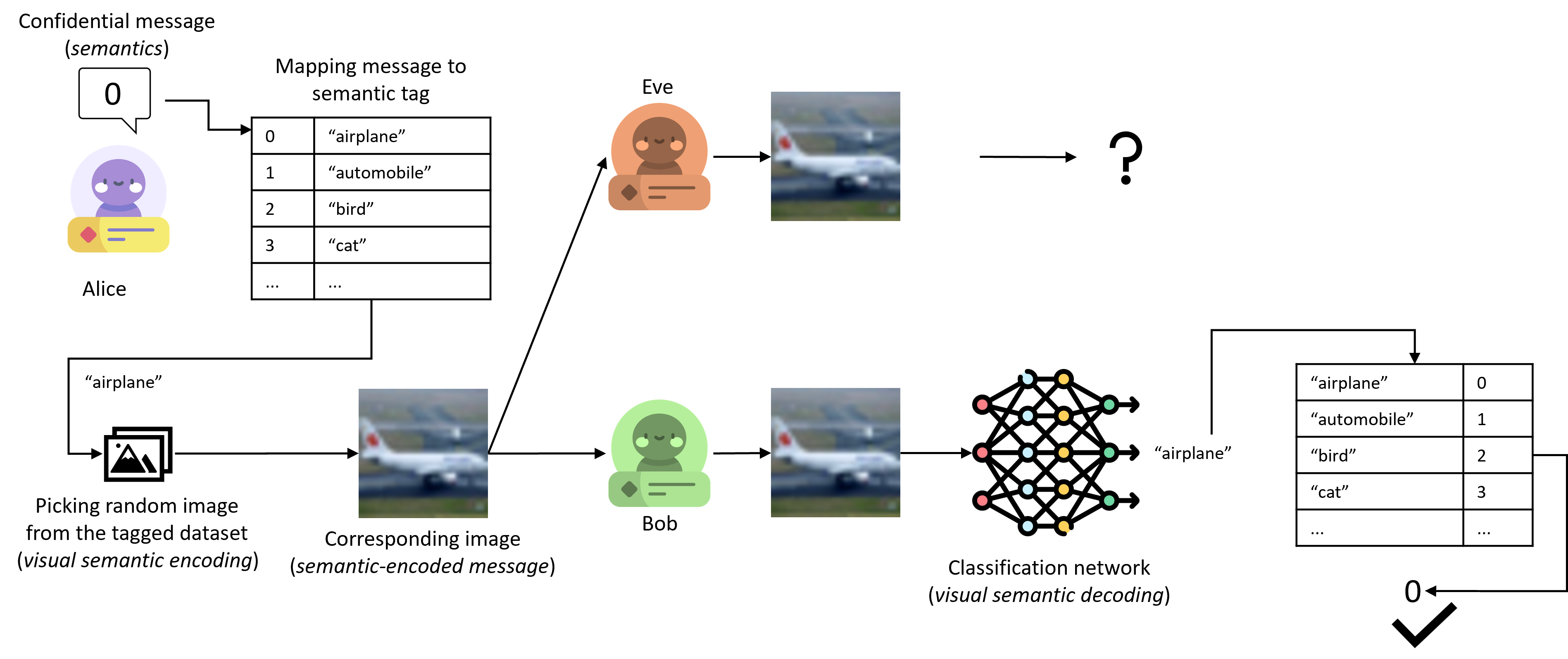}
        \caption{{Naive approach: \emph{Alice} maps messages to semantic tags and transmits corresponding images. \emph{Bob} uses a classification network to decode. Security relies entirely on keeping the semantic mapping table secret from \emph{Eve}, failing when \emph{Eve} is an insider or protocols are standardized.}}
        \label{subfig:naive_design}
    \end{subfigure}
    \vspace{1mm}
    \hrule
    \vspace{1mm}
    \begin{subfigure}[t]{\linewidth}
        \centering
        \includegraphics[scale=.55]{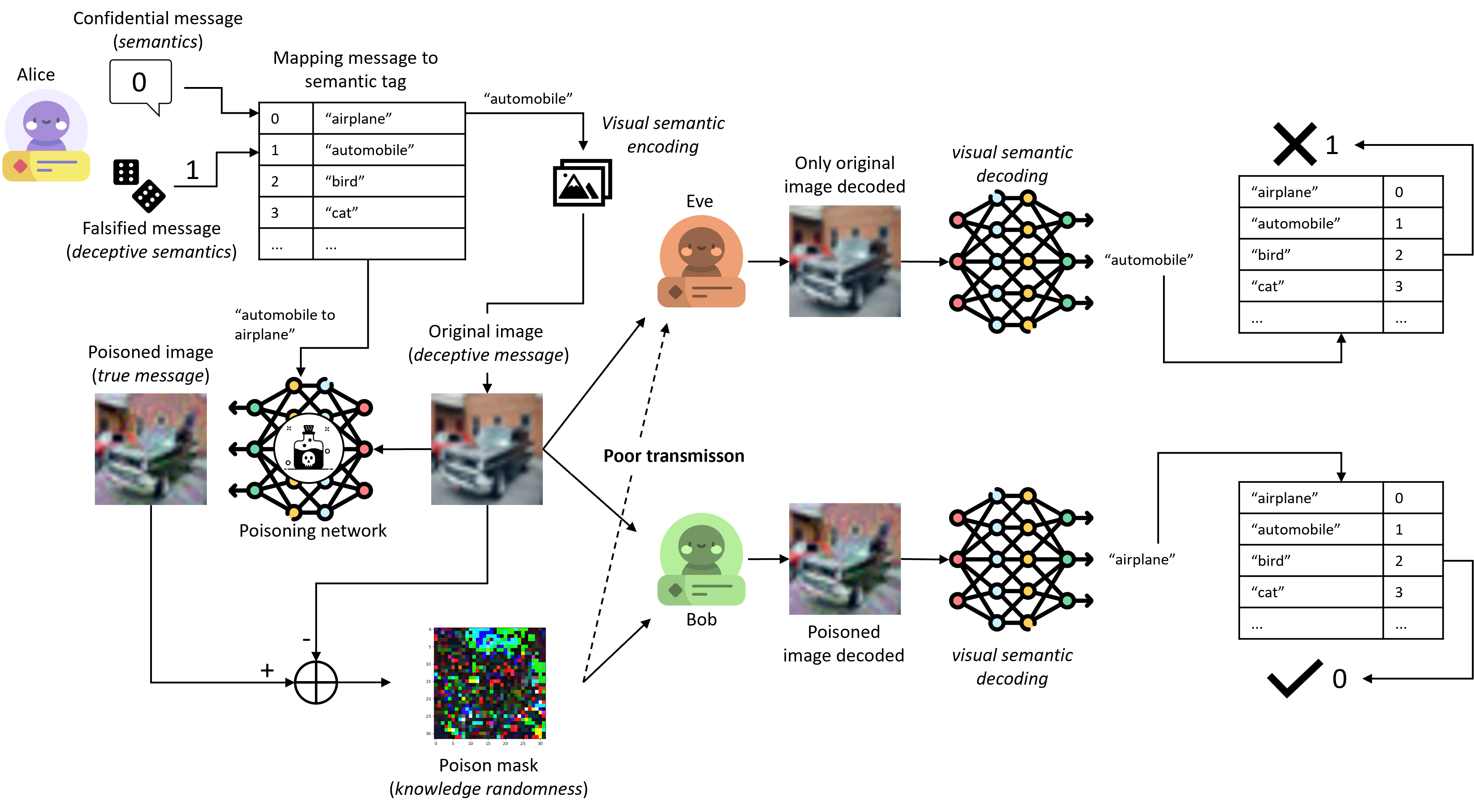}
        \caption{{\ac{venena}: \emph{Alice} generates a falsified message encoded in an original image (deceptive content), then creates a poisoned version that classifies as the true message. The poison mask (difference between images) and original image are power-multiplexed and transmitted. \emph{Bob} decodes both components with favorable channel conditions and reconstructs the poisoned image to recover the true message. \emph{Eve} decodes only the high-power original image due to worse channel conditions, perceiving the falsified message. \ac{venena} remains secure even with full system knowledge, as the poison mask provides semantic confidentiality.}}
        \label{subfig:venena}
    \end{subfigure}
    \caption{Visual encryption of confidential messages: \subref{subfig:naive_design} naive design to defend eavesdroppers without essential knowledge for decryption, and \subref{subfig:venena} the \ac{venena} framework to deceive insider eavesdroppers.}
    \label{fig:design}
\end{figure*}

\subsection{VISUAL ENCRYPTION}\label{subsec:visual_encryption}
To realize such confidential transmission, \emph{Alice} 
% can map every raw message, which can be {considered} as the semantics to deliver, onto a semantic tag (e.g., ``{airplane}'', ``automobile'', or ``bird''), to which an image dataset is associated. Instead of directly sending the raw message, \emph{Alice} randomly selects an image from the subset of images with the corresponding semantic tag, and {sends} it to \emph{Bob}. Thus, the raw message (number) is visual-semantically encoded into an visual message (image). \emph{Bob} will then estimate the semantic tag associated to the received image with a pre-trained classification network. 
{employs a semantic codec that operates through categorical labels called semantic tags (e.g., ``airplane'', ``automobile'', or ``bird''). Each raw message is mapped onto a semantic tag, to which an image dataset is associated. The encoding process randomly selects an image from the corresponding dataset, transforming the raw message (number) into a visual message (image). \emph{Bob} then uses a pre-trained classification network to extract the semantic tag and recover the message.}
As long as the mapping between the numbers and the semantic tags is unknown to \emph{Eve}, the semantic information remains confidential even if \emph{Eve} can correctly decode the image. This visual encryption approach is illustrated in Fig.~\ref{subfig:naive_design}.

% Both the legitimate and eavesdropping channels are lossy, yet with sufficiently high capacity to transmit a image at required data rate.

Obviously, the security of the visual encryption scheme completely relies on the secrecy of the mapping between the message codewords and the semantic tags. Unfortunately, there are multiple cases where the confidentiality of this information can be compromised. For example, when the protocol design is standardized and open, or when the eavesdropper is a malicious insider, or when the eavesdropper is another legitimate receiver of the system, it becomes possible to learn the mapping from statistical analysis of the eavesdropped images. In such cases, the visual encryption scheme fails to provide the desired information secrecy.

% For a worst-case analysis, in this study we consider \emph{Eve} as an ``insider'' eavesdropper who has full knowledge about the system design, including the security mechanism, the protocol, the data format, the parameter specifications, and the image classification network used by \emph{Bob}. In other words, \emph{Bob} and \emph{Eve} share the identical knowledge.

\subsection{DECEPTIVE VISUAL ENCRYPTION}\label{subsec:deceptive_visual_encryption}
To enhance the security of the visual encryption scheme, we propose a novel framework called \ac{venena}, which leverages the concept of \ac{pld} and image poisoning. As shown in Fig.\ref{subfig:venena}, instead of simply sending an image with the message-associated semantic tag, \emph{Alice} first randomly generates a falsified message, which is then visual-encoded into a correspondingly tagged image, which we call the \emph{original image}. This image serves as a deceptive message, since it carries deceptive semantics. It is then processed with a poisoning network in such a way, that the \emph{poisoned image} at output shall be classified with the true semantic tag, while remaining miminally diversed from the original image in the sense of error power. It is this poisoned image that carries the true semantics. Afterwards, a \emph{poison mask} is obtained by pixel-wise differentiating the original image from the poisoned one. Seen from the semantic \ac{pld} perspective, the poison mask carries a random \revthree{knowledge} that maintains the confidentiality of the true semantics, when both the semantic codec and the deceptive message are leaked~\cite{HZS+2025semantic}. The original image and the poison mask are then multiplexed and sent together,
%with the resource allocation between them fine tuned so that the poison mask component has only slightly higher \ac{sinr} at \emph{Bob} than required for successful decoding, while  the original image component has sufficient \ac{sinr} at \emph{Bob} for a significant redundancy in channel capacity. 
{supposed to be decoded by \emph{Bob} separately for construction of the poisoned image. Specifically, while \ac{venena} is agnostic to the multiplexing method, in this work we apply a power-domain non-orthogonal multiplexing, where the original image is allocated with significantly higher power than the poison mask, so that it can be first decoded treating the poison mask as a weak inteference. \revthree{This introduction of inteference, of course, will reduce the \ac{sinr} of the received original image and therewith the classification accuracy for both \emph{Bob} and \emph{Eve}.} Subsequently, the decoded original image is subtracted from the received signal, and the poison mask can be decoded from the residual signal (known as the technique of \ac{sic}~\cite{SSC+2013successive}). The power allocation shall be fine tuned in such way, that the \ac{sinr} at \emph{Bob} \begin{enumerate*}[label=(\roman*)]
    \item significantly exceeds the necessary level for successful decoding of the original image, but
    \item fulfills the requirement for decoding of the poison mask with only a small margin.
\end{enumerate*}
}
Thus, \ac{pls} is selectively applied on the poison mask component, which carries the differential information between the true and falsified messages, while the original image carrying falsified message is well exposed to potential eavesdroppers.

With appropriate beamforming by \emph{Alice} towards \emph{Bob}, combined with other common measures such like secrecy protected zones, it is generally agreed that \emph{Eve} in wireless eavesdropping scenarios suffers from statistically worse channel conditions than \emph{Bob} does, as explained in~\cite{CHZ+2025physical}. Therefore, while the original image, due to its high power, may be well decoded by both \emph{Bob} and \emph{Eve}, the poison mask is only decodable by \emph{Bob}. Having both components decoded, \emph{Bob} will be able to regenerate the poisoned image, and therewith obtain the true message. In contrast, missing the poison mask, \emph{Eve} will only perceive the falsified message carried by the original image, and therefore be deceived.

% the concept of \ac{pld} with image poisoning. The \ac{venena} framework aims to deceive eavesdroppers, especially insider eavesdroppers, by introducing artificial noise into the transmitted images. This noise is carefully crafted to mislead the eavesdropper into making incorrect inferences about the transmitted message, even if the eavesdropper has full knowledge of the system design.

\section{IMAGE POISONING}\label{sec:poisoning}
% \todo[bh,inline]{Review and revise}

While the \ac{venena} framework provides the overall approach for deceptive visual encryption, its effectiveness heavily relies on the image poisoning technique that enables semantic transformation while maintaining visual similarity. In this section, we present a novel targeted data poisoning attack specifically designed for this purpose.
% \subsection{Poisoner}
% In this section, we present a novel targeted data poisoning attack in the context of a sender, receiver, and eavesdropper scenario. The proposed method converts the clear image (class A) into a poisoned one by target class B that matches the gradients of the target example.

% \subsubsection{Threat Model}
% We consider a setting with three parties: the sender, the receiver, and the eavesdropper. The sender has access to a clean image $C$ and produce a poison $P$. The sender can poison by perturbing images within an $\ell_{\infty}$ $\epsilon$-bound (e.g., $\epsilon \leq 16$). The sender has full control over the classification model and the poisoning process. Finally, the sender can provide the clean image $C$ and corresponding poison $P$. For example, if the sender wants to send information belonging to class 3, it will send a clean image (except class 3) and poison which makes the image change to class 3.

 {Given a clear image $I_A$ belonging to the class $A$ and a target class $B$, the desired method is supposed to poison $I_A$ to $I_B^\text{p}$ by perturbing images a fraction of images. This poison process is accomplished by \ac{gma}, which exploits gradient descent learning: during training, each sample contributes a gradient guiding the model's parameter updates. \ac{gma} crafts poisoned samples whose training gradients align with the adversarial gradient that would cause the target misclassification ($A\to B$). By aligning these gradient directions, the poisoner steers the model's learning process toward the desired confusion, such that the trained model incorrectly predicts class B for the target image $I_A$ during inference. Critically, perturbations are constrained to remain visually imperceptible (bounded by $\epsilon$), ensuring the attack remains stealthy.}

%\todo[bh,inline]{Is it necessary to capitalize the term gradient matching attack here (coz you stop doing that in the subsequent text.) The same for the title of Algorithm 1}

% The receiver receives the clean image $C$, poison $P$ and the trained model $F_{\theta}$ from the sender. The receiver has knowledge of the gradient matching attack and can fusion the poison image $P$ and the clean image $C$ to get the target information.

% The eavesdropper intercepts the clean image $C$ sent by the sender, but was unable to eavesdrop on the poison $P$. Due to the distortion applied by the sender, the eavesdropper cannot restore the target information from the clean image $C$ and the poison image $P$.

% \subsection{Gradient Matching Attack}
\begin{algorithm}[!htbp]
% \captionsetup{font={color=\highlightcolor}}
\caption{{The \ac{gma} algorithm}}
\label{alg:gma}
\begin{algorithmic}
% \color{\highlightcolor}
\Require Clean image $I_A$, Target class $T$, Poison budget $\epsilon$, Loss function $\mathcal{L}$, Trained model $F_{\theta}$, Hyperparameters $\lambda_1, \lambda_2, \lambda_3$
\Ensure Poison $P$
\State Initialize $P \gets C$
\For{$i = 1, \ldots, n_\text{iter}$}
\State Sample a clean batch ${(x_c, y_c)} \sim D_c$
\State $\theta \gets \theta - \eta \nabla_{\theta} \mathcal{L}(F_{\theta}(x_c), y_c)$
\State $g_t \gets \nabla_{\theta} \mathcal{L}(F_{\theta}(T), y_t)$
\State $g_p \gets \nabla_{\theta} \mathcal{L}(F_{\theta}(P), y_p)$
\State $\mathcal{L}_\text{match} \gets -\sum{l=1}^L \frac{g_t^l \cdot g_p^l}{|g_t^l|2 |g_p^l|2}$
\State $\mathcal{L}_\text{class} \gets \text{CrossEntropy}(F{\theta}(P), y_t)$
\State $\mathcal{L}_\text{norm} \gets \sum{l=1}^L |g_p^l|2^2$
\State $\mathcal{L}_\text{total} \gets \lambda_1 \mathcal{L}_\text{match} + \lambda_2 \mathcal{L}_\text{class} + \lambda_3 \mathcal{L}_\text{norm}$
\State $P' \gets P - \alpha \nabla{P} \mathcal{L}_\text{total}$
\State $P' \gets \text{clip}(P', P - \epsilon, P + \epsilon)$
\State $P \gets P'$
\EndFor
\State Train the classification model $F{\theta}$ on $C \cup P$
\State \Return $P$, $F_{\theta}$
\end{algorithmic}
\end{algorithm}

% The process of poisoner aims to cause misclassification of a clean image $x^c$ by creating poison data whose training gradient correlates with the adversarial target gradient. This is achieved by aligning the target and poison gradients in the same direction through minimizing their negative cosine similarity:

% \begin{align}
% g_{t} &= \nabla_{\theta} L(F(x^t, \theta), y^{adv}) \\
% g_{p} &= \sum_{i=1}^{P} \nabla_{\theta} L(F(x_i + \Delta_i, \theta), y_i) \\
% B(\Delta, \theta) &= 1 - \frac{\langle g_{t}, g_{p} \rangle}
% {\left| g_{t} \right| \cdot \left| g_{p} \right|}
% \end{align}

% \begin{equation}
% B(\Delta, \theta) = 1 - \frac{\langle\nabla_{\theta}L(F(x^t, \theta), y^{adv}), \sum_{i=1}^{P} \nabla_{\theta}L(F(x_i + \Delta_i, \theta), y_i)\rangle}
% {\left|\nabla_{\theta}L(F(x^t, \theta), y^{adv})\right| \cdot \left| \sum_{i=1}^{P} \nabla_{\theta}L(F(x_i + \Delta_i, \theta), y_i)\right|}
% \end{equation}

% where $\Delta$ represents the perturbations applied to the poison examples, $\theta$ represents the model parameters, $L$ is the loss function, $F$ is the model, $x^t$ is the target example, $y^{adv}$ is the adversarial target label, $x_i$ are the poison examples, and $y_i$ are the corresponding labels.

To achieve this, the poisoner computes two sets of gradients: the target gradient and the poison gradient. The target gradient represents the direction of change in the model's parameters that would increase the loss for the target image with respect to the adversarial target label. The poison gradient, on the other hand, is the sum of the gradients of the poisoned examples with their corresponding labels.

The poisoner then minimizes the negative cosine similarity between the target gradient and the poison gradient. By aligning these gradients in the same direction, the poisoner ensures that the model's updates during training will move in a direction that increases the likelihood of misclassifying the target image.

By iteratively updating the perturbations to minimize the negative cosine similarity between the target and poison gradients, the poisoner crafts poisoned examples that effectively manipulate the model's behavior, causing it to misclassify the target image during inference. Throughout this optimization process, the perturbations are constrained within $\epsilon$. {A pseudo code of \ac{gma} is presented in Algorithm~\ref{alg:gma}.}

\section{NUMERICAL VERIFICATION}\label{sec:simulation}

\subsection{SIMULATION SETUP}\label{subsec:setup}
\subsubsection{Data Format}
Every confidential message is a random digit in $\mathcal{M}=\{0,1,2,\dots 9\}$. An image database containing $10$ classes of photos{, generated from the CIFAR-10 data,} is therefore used for the visual encryption, the mapping between messages and image classes (tags) is listed in Tab.~\ref{tab:mapping}. Every {message image} is $32\times 32$ pixels {in} RGB, with each pixel quantized to $8$ bits per channel.

\begin{table}
    \centering
    \caption{Mapping between messages and semantic tags}
    \label{tab:mapping}
    \begin{tabular}{c | >{\centering\arraybackslash}p{1cm} >{\centering\arraybackslash}p{1cm} >{\centering\arraybackslash}p{1cm} >{\centering\arraybackslash}p{1cm} >{\centering\arraybackslash}p{1cm}}
        \toprule[2px]
        \textbf{Message} & 0 & 1 & 2 & 3 & 4\\
        \textbf{Tag} & airplane & automobile & bird & cat & deer\\
        \midrule[1.5px]
        \textbf{Message} & 5 & 6 & 7 & 8 & 9\\
        \textbf{Tag} & dog & frog & horse & ship & truck\\
        \bottomrule[2px]
    \end{tabular}
\end{table}

\subsubsection{{Networks Training and Evaluation}}\label{subsubsec:networks_training}
\paragraph*{\textbf{Classifier Network}} {Since \ac{gma} is designed to deceive specific neural networks, we must first train the implementation of the classifier network on the receiver's side. It} employs a pre-trained {v}ision {t}ransformer architecture, trained offline on the CIFAR-10 dataset without poisoned data. This pre-training approach ensures that during operational deployment, the system requires only inference operations with minimal computational cost. The victim network is trained using the specified hyperparameters and configuration options. The training image size is set to $224\times 224$ pixels, and the batch sizes for training and evaluation are both set to 64. 

The learning rate for the victim network is set to 0.03 with a cosine decay schedule. The weight decay is set to 0, and the total number of training steps is 10~000. The learning rate warmup is performed for the first 500 steps, and the maximum gradient norm is clipped to 1.0.
The victim network training utilizes mixed-precision training (FP16) with an optimization level of O2. 

{Table \ref{tab:time_performance} presents the inference time evaluation of our classifier model. With $L\cdot d=11.17$ million parameters (memory size \SI{42.63}{\mega\byte})}, the model achieves a mean inference time of ${1.8866\pm 0.2258}\si{\milli\second}$ per image, yielding a throughput of 530.04 images per second. The low standard deviation (12\% of the mean) and stable min-max range (${1.69\--3.19}\si{\milli\second}$) demonstrate consistent performance suitable for real-time applications, making it practically viable {even} for real-time \ac{6g} communication scenarios.

\begin{table}[htbp]
\centering
\caption{{Testbed setup and real-time performance}}
\label{tab:time_performance}
\begin{tabular*}{\linewidth}{@{\extracolsep{\fill}}lc@{}}
\toprule[2px]
\textbf{Metric} & \textbf{Value} \\
\midrule[1px]
\multicolumn{2}{c}{{Testbed Setup}} \\
{Classifier Network Parameters (\#)} & {11~173~962} \\
{Classifier Network  Size (MB)} & 42.63 \\
GPU & NVIDIA GeForce RTX 4090 \\
\midrule[1px]
\multicolumn{2}{c}{{\textit{Poisoning Time}}}\\
{Mean Time per Image (s)} & {0.3170}\\
{Standard Deviation (s)} & {0.0045}\\
{Min Time per Image (s)} & {0.3080}\\
{\revthree{Max} Time per Image (s)} & {0.3260}\\
{Throughput (images / s)} & {3.15}\\
\midrule[1px]
\multicolumn{2}{c}{{\textit{Inference Time}}}\\
{Mean Time per Image (ms)} & {1.8866}\\
{Standard Deviation (ms)} & {0.2258}\\
{Min Time per Image (ms)} & 1.6900 \\
{\revthree{Max} per Image (ms)} & 3.1937 \\
{Throughput (images / s)} & 530.04 \\
\bottomrule[2px]
\end{tabular*}
% \end{tcolorbox}
\end{table}

\paragraph*{\textbf{Poisoning Process}} 
% The data poisoning process employs a gradient matching attack that is executed offline before deployment, eliminating real-time computational overhead during transmission. The attack utilizes an ensemble of surrogate models that are trained on clean CIFAR-10 data. During the offline poisoning phase, these surrogate models guide an iterative optimization process that generates adversarial perturbations for 10\% of samples from the source class. The optimization aligns the gradients of poisoned samples with those that would cause misclassification of target images into the source class, while constraining perturbations within an $L_\infty$ bound of $\epsilon = 16/255$. Once the poison perturbations are computed offline, they are stored and can be applied to images during transmission with negligible computational cost—requiring merely a few microseconds per image for adding the pre-computed perturbations. 

% Once the poison perturbations are computed offline, they are stored and can be applied to images during transmission with negligible computational cost—requiring merely a few microseconds per image for adding the pre-computed perturbations. 

{The data poisoning process employs a gradient matching attack, which utilizes an ensemble of surrogate models that are offline trained on clean CIFAR-10 data. During the poisoning phase, these surrogate models guide an iterative optimization process that generates adversarial perturbations for 10\% of samples from the source class (the penetration rate is empirically validated). The optimization aligns the gradients of poisoned samples with those that would cause misclassification of target images into the source class, while constraining perturbations within an $L_\infty$ bound of $\epsilon = 16/255$. 

Each poisoning operation requires $n_\text{iter}$ iterations of gradient computations across all $L$ network layers, resulting in $O(n_\text{iter} \cdot L \cdot d)$ complexity where $d$ represents parameter dimensionality. The layer-wise gradient matching $\mathcal{L}_\text{match} = -\sum_{l=1}^L \frac{g_t^l \cdot g_p^l}{\left\vert g_t^l\right\vert_2 \left\vert g_p^l\right\vert_2}$ requires substantial memory for gradient tensor storage and manipulation. 

To assess the time efficiency of the poisoning process, we carried out the poisoning on $5000$ image samples w.r.t. the classifier network trained above, with $n_\text{iter}=500$ iterations, and measuring the poisoning time per image. The testbed setup and test results are summarized in Table~\ref{tab:time_performance}. As we can see, the poisoning process performs at a stationary speed with average throughput of $3.15$ images per second, which can be sufficient for low-rate transmissions of sparse confidential messages, but far from real-time for high-rate data streams.

% Therefore, we adopt an offline poisoning strategy using pre-trained surrogate models on clean CIFAR-10 data. During the offline phase, Algorithm~\ref{alg:gma} executes without time constraints to train poisoned models $F_{\theta}$ for various source-target class pairs, optimizing the poisoning process within the $L_\infty$ bound of $\epsilon = 16/255$. In the online phase, the system only needs to perform forward inference through the pre-trained poisoned classifier, reducing computational requirements from expensive iterative optimization to a single classification inference pass. This transforms the online complexity from $O(n_{iter} \cdot L \cdot d)$ gradient operations to $O(L)$ forward propagation, enabling real-time deployment of the VENENA framework while maintaining the security properties of gradient matching.

Shall \ac{venena} be applied in scenarios where high data rates are required, we recommend to rely on offline-poisoned image databases instead of carrying out poisoning online. With poisoned images pre-generated and stored for immediate use during transmission, reducing the time complexity from $O(n_\text{iter} \cdot L \cdot d)$ to $O(1)$ per image. An \revthree{occasional} update of the poisoned image database shall be scheduled in this case, to eliminate long-term statistical learning by eavesdroppers.}

\subsubsection{Transmission Scheme}\label{subsubsec:transmission_scheme}
For every message $m\in\mathcal{M}$, which is one digit between 0 and 9, a distinct digit $\tilde m\neq m$ is randomly selected as the fake message. An image $i_m$ is then randomly picked from the data base $\mathcal{I}_m$ that corresponds to the semantic tag associated with $m$. This image is then poisoned by the network $\mathcal{P}_m^{\tilde m}$ to generate its $\tilde m$-induced version $i_m^{\tilde m}$. Both images are interpreted in the format of binary vectors and then their sum modulo two is calculated to obtain the binary vector of additive poison 
{
\begin{equation}
    \left(p_m^{\tilde m}\right)\subscript{b}=\left(i_m\right)\subscript{b} \mathrm{XOR} \left(i_m^{\tilde m}\right)\subscript{b}.
\end{equation}}
Both $i_m$ and $p_m^{\tilde m}$ are then modulated in \ac{bpsk} into the waveform components $s_i$ and $s_p$, respectively. The two components are then mixed 
%in the power domain with a mixing ratio $\alpha$ to form the transmitted signal $s(t)=\sqrt{\alpha}s_i(t)+\sqrt{1-\alpha}s_p(t)$, and $s$ is then transmitted with power of $P_\Sigma$.
together with a power ratio $\alpha:(1-\alpha)$ into the transmitted signal $s$, which is transmitted at the power $P_\Sigma${, i.e., 
\begin{equation}
    s(t)=\sqrt{\alpha P_\Sigma}s_i(t)+\sqrt{(1-\alpha) P_\Sigma}s_p(t).
\end{equation}
}

\subsubsection{Eavesdropper Setup}\label{subsubsec:eavesdropper_setup}
We consider two different knowledge levels for \emph{Eve}: \begin{enumerate}
    \item \emph{Full knowledge}, where \emph{Eve} shares the identical knowledge with \emph{Bob} about the system design, including the \revise{\ac{venena} framework}, the protocol, the data format, the parameter specifications, the image classification network used by \emph{Bob}, \revise{as well as the mapping table}\revthree{. In this case, \emph{Eve} will try to carry out \ac{sic} to decode both the original image and the poison mask, and then reconstruct the poisoned image for message perception. If it fails to decode the poison mask, it will assume that \ac{venena} is not activated, and perceive the message carried by the original image only.}
    \item \emph{Partia knowledge}, where \emph{Eve} knows the basic visual encryption scheme and the mapping table, but is not aware of the \ac{venena} framework.
\end{enumerate}
% \revise{In this article, we focus on the case where \emph{Eve} to be consistently eavesdropping. For readers who are interested in sophisticated dynamic eavesdropping strategies, we refer to our related work~\cite{HZS+2025semantic,CHZ+2025physical3}.}
\subsubsection{Radio Setup}\label{subsubsec:radio_setup}
We set the channel bandwidth to \SI{1}{\mega\hertz} and the gross transmission bit rate to \SI{1}{\mega\bit}/\si{\second}. \revise{To reflect realistic multipath wireless scenarios, we consider Rayleigh fading \ac{nlos} channels for both \emph{Bob} and \emph{Eve}, with a higher mean path loss in the eavesdropper channel than that in the legitimate channel.} The power density of \ac{awgn} is \SI{-174}{\dBW}/\si{\hertz} at both receivers.

\subsection{VALIDATION TEST}\label{subsec:validation}
\begin{table*}[!htbp]
    \centering
    % \begin{tcolorbox}[colframe=blue, colback=white, sharp corners, halign=center]
    \caption{Benchmark results: message perception accuracy}
    \label{tab:benchmarks}
    \begin{tabular}{l c c c c}
        \toprule[2px]
         & {\textbf{VENENA, 75\% mixing}} & {\textbf{NVE, full power}} & {\textbf{NVE, 75\% power}}& {\textbf{NVE, conventional PLS*}}\\
         \midrule[1.5px]
         \textbf{Bob}                       & 93.43\% & 97.22\% & 96.91\% & 93.42\%\\
         \textbf{Eve (full knowledge)}      & 51.12\% & 88.96\% & 86.82\% & 60.22\%\\
         \textbf{Eve (partial knowledge)}   & 5.18\%  & 89.14\% & 86.66\% & 60.22\%\\
         \bottomrule[2px]
    \end{tabular}
    % \end{tcolorbox}
\end{table*}

\begin{figure*}[!htbp]
    \centering
    \includegraphics[width=.7\linewidth]{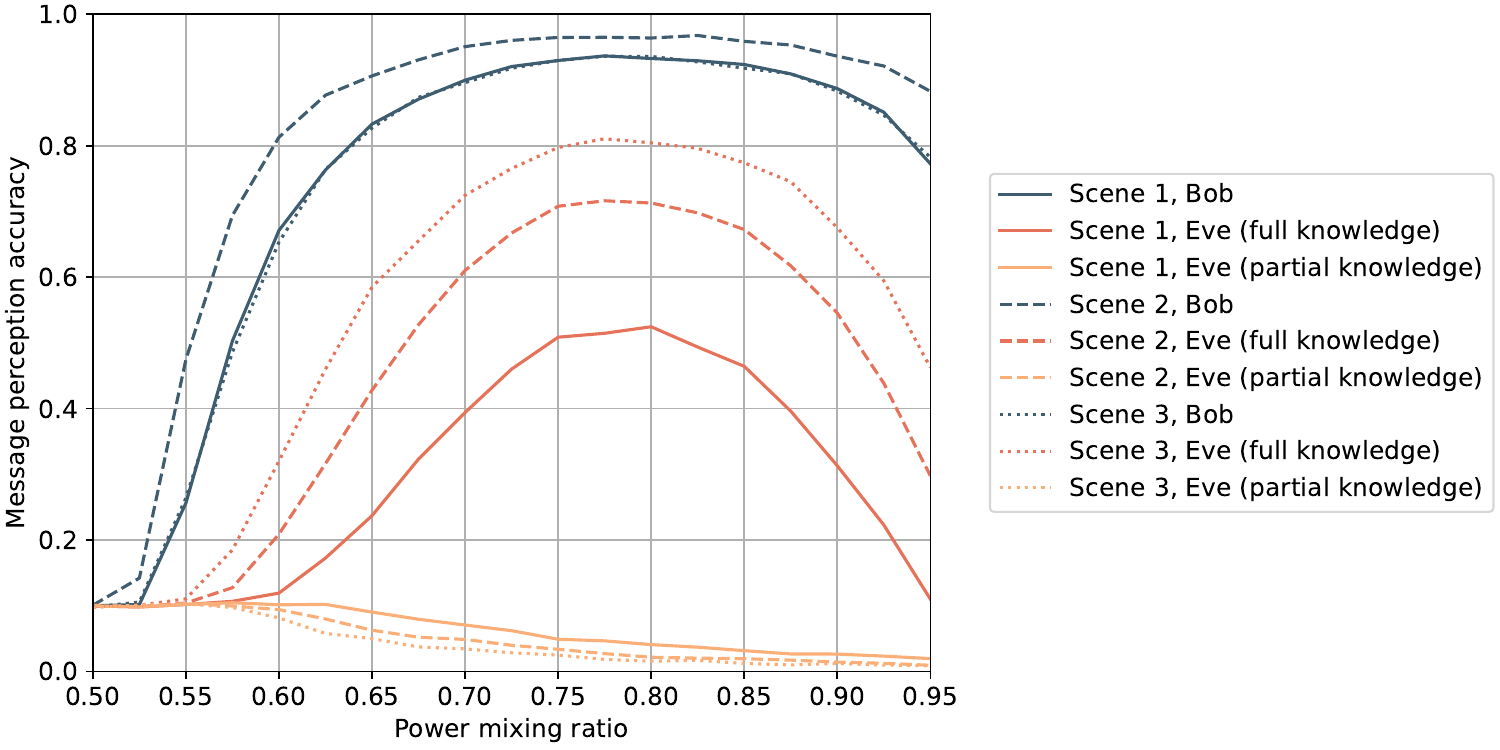}
    \caption{{Sensitivity of message perception accuracy versus mixing ratio $\alpha$ under different transmission power and eavesdropping channel conditions.}}
    \label{fig:sensitivity_test}
\end{figure*}

First we conducted a validation test to demonstrate the capability of the \ac{venena} framework to securely transmit confidential messages. We considered the transmission power budget $P_\Sigma=\SI{100}{\milli\watt}$, \revise{mean legitimate channel gain $\overline{z}\subscript{{Bob}}=\SI{-85}{\dB}$, mean eavesdropping channel gain $\overline{z}\subscript{Eve}=\SI{-95}{\dB}$,} and the following \revise{four} schemes for benchmarking:
\begin{enumerate}
    \item \emph{\ac{venena}, 75\% mixing}: $\alpha=0.75$, $P_\Sigma=\SI{100}{\milli\watt}${, which provides \emph{Bob} a small \ac{sinr} of \SI{3}{\dB} for the poison mask decoding, corresponding to channel capacity of \SI[per-mode = symbol]{1.58}{\mega\hertz\per\second} that slightly exceeds the transmission rate};
    \item \emph{\ac{nve}, full power}: $\alpha=1$, $P_\Sigma=\SI{100}{\milli\watt}$;
    \item \emph{\ac{nve}, 75\% power}: $\alpha=1$, $P_\Sigma=\SI{75}{\milli\watt}$; and
    \item \emph{\ac{nve}, conventional \ac{pls} baseline}: $\alpha=1$, transmission power adjusted to the level that \emph{Bob} achieves the same perception accuracy as in the \ac{venena}-75\%-mixing scheme.
\end{enumerate}

We carried out the Monte-Carlo test with $10~000$ independent trails under each scheme, where \emph{Bob}, full-knowledge \emph{Eve}, and partial-knowledge \emph{Eve} are measured regarding their and capability of accurately perceiving the confidential message. The results are summarized in Tab.~\ref{tab:benchmarks}. It can be observed that \revise{compared to all \ac{nve} baselines including that assisted by conventional \ac{pls},} the \ac{venena} framework can effectively deceive \emph{Eve} even when it has full knowledge about the system design, while maintaining a high message perception accuracy for \emph{Bob}.

{These optimization objectives can be realized through dynamic mixing strategies in practical deployment. The optimal mixing ratio $\alpha$ depends on both legitimate and eavesdropping channel conditions. While \emph{Bob}'s channel condition can be directly measured through standard channel estimation techniques, \emph{Eve}'s channel statistics can be estimated based on geographical security zones, historical eavesdropping patterns, or worst-case assumptions for conservative design. A practical implementation would pre-compute performance lookup tables through large-scale simulation campaigns covering diverse channel condition combinations. During operation, the mixing ratio is adaptively selected to maximize the chosen objective (e.g., accuracy gap or \emph{Eve}'s failure rate) while maintaining \emph{Bob}'s minimum accuracy requirement. This channel-sensing-based dynamic mixing enables \ac{venena} to adapt to time-varying wireless environments.}

\subsection{SENSITIVITY TEST}\label{subsec:sensitivity}
We further conducted a sensitivity test to evaluate the sensitivity of the performance of \ac{venena} against the transmission power and the eavesdropping channel condition under different mixing ratios. Similarly to the verification test, we considered \emph{Bob} with \revise{mean channel gain $\overline{z}\subscript{Bob}=\SI{-85}{\dB}$}, \emph{Eve} with full knowledge, and \emph{Eve} with partial knowledge, and investigate their performance in three different scenes:
\begin{enumerate}
    \item \emph{Scene 1}: $P_\Sigma=\SI{100}{\milli\watt}$, $\revise{\overline{z}\subscript{Eve}=\SI{-95}{\dB}}$;
    \item \emph{Scene 2}: $P_\Sigma=\SI{200}{\milli\watt}$, $\revise{\overline{z}\subscript{Eve}=\SI{-95}{\dB}}$; and
    \item \emph{Scene 3}: $P_\Sigma=\SI{100}{\milli\watt}$, $\revise{\overline{z}\subscript{Eve}=\SI{-90}{\dB}}$.
\end{enumerate}
In all three scenes we measured the message perception accuracy of \emph{Bob} and \emph{Eve} under different mixing ratios ranging from $0.5$ to $0.95$. For each scene and each mixing ratio, we conducted $1000$ independent trails.

The results are depicted in Fig.~\ref{fig:sensitivity_test}, from which we can observe that \emph{Eve}'s performance monotonically decreases with the raising mixing ratio $\alpha$ in all three scenes, as long as it is not aware of the \ac{venena} framework. For both \emph{Bob} and \emph{Eve} with full knowledge, the performance is generally concave of the power mixing ratio in all three scenes, with the maximum perception accuracy achieved around $75\%$ mixing. The optimal specification of the overall system, however, depends on the optimaziation objective, e.g. the perception accuracy gap between \emph{Bob} and \emph{Eve}, or the perception failure rate of \emph{Eve} under the constraint of \emph{Bob}'s mimimum perception accuracy.

\begin{figure}[!htbp]
    % \captionsetup{font={color=\highlightcolor}}
    \centering
    \revboxtwo{\includegraphics[width=.95\linewidth]{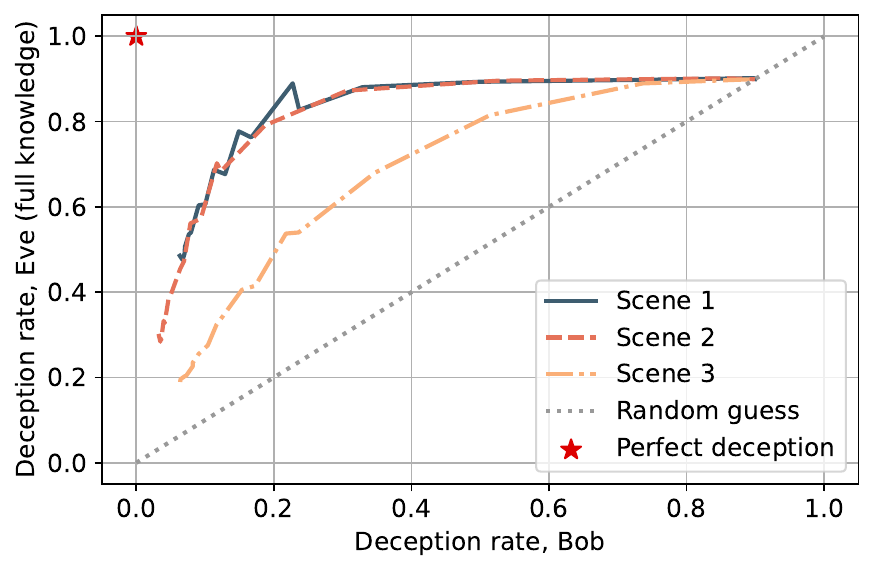}}
    \caption{Deception rates of \emph{Bob} and full-knowledge \emph{Eve}, respectively, under different mixing ratios in three scenes.}
    \label{fig:roc}
\end{figure}

{For a more intuitive illustration of \ac{venena}'s descriminative deception on \emph{Eve}, we present in Fig.~\ref{fig:roc} the \ac{roc} curves showing the deception rate of \emph{Bob} and full-knowledged \emph{Eve} upon different mixing ratios in the three scenes. }

\section{DISCUSSION}\label{sec:discussion}
Towards a more practical deployment of the \ac{venena} framework in realistic communication scenarios, some discussions regarding the efficiency aspects are necessary.
\subsection{ENCRYPTION CODING EFFICIENCY}\label{subsec:coding_efficiency}
The encryption coding efficiency of the \ac{venena} framework is determined by the size of the image and the amount of available semantic tags. For the demonstrational setup we have used in this article, the coding rate is as low as $\log_2^{10} / (32\times 32\times 3\times 8)\approx 1.35\times 10^{-4}$, which is applicable only for low-speed communication with high secrecy requirements. To raise the efficiency for high-speed data transmission scenarios, {several approaches can be considered:
\begin{enumerate}
    \item \textbf{Smaller images}: Reducing image dimensions directly decreases transmission overhead, though this will degrade the decoding accuracy of the classification network.
    \item \textbf{More semantic tags}: Introducing additional semantic tags to the mapping table increases coding rate logarithmically, but requires a larger image database and more poisoning networks to be trained.
    \item \textbf{Multi-tag images}: Using images containing multiple semantic tags simultaneously (e.g., an image showing both an airplane and a bird) significantly increases information density. However, this substantially complicates both database construction--requiring systematic labeling of multi-tag combinations--and poisoning network training, as the network must achieve targeted misclassification across multiple semantic dimensions simultaneously.
    \item \textbf{Multi-modal encoding}: Extending beyond visual content to incorporate audio or text modalities enables parallel transmission of multiple semantic channels, multiplicatively increasing the effective coding rate. This requires developing poisoning networks for each modality and ensuring cross-modal consistency.
    \item \textbf{Semantic-preserving compression}: Pre-processing images with neural compression techniques can reduce transmission overhead significantly while maintaining classification accuracy, achieving superior rate-distortion tradeoffs compared to traditional codecs without requiring additional database or network resources. It shall be noted that this approach introduces additional computational complexity and delay at both the transmitter and receiver.
\end{enumerate}
\revthree{The specific implementation of these approaches, however, are out of the scope of this article, and shall be left for future works.}
}

\subsection{COMPLEXITY AND SCALABILITY}\label{subsec:scalability}
By increasing the number of semantic tags, the encryption coding efficiency can be linearly improved. However, this will also raise the space complexity, as it also requires linear increases in both the size of image database and the set of poisoning networks (since each network is specified to a dedicated semantic pair). {Although the complexities of runtime encoding, poisoning, transmission, and decoding are not impacted by the semantic tag number $n$, a $\mathcal{O}(n)$ space complexity in database, a $\mathcal{O}(n^2)$ space complexity in poisoning networks, and a $\mathcal{O}(n^2)$ time complexity in network training are expected. Advanced techniques such as data distillation~\cite{YLW2024dataset} and network pruning~\cite{CZS2024survey} shall be considered, in this context, to improve the scalability of the \ac{venena} framework for practical deployment with a large number of semantic tags.
}

{
\subsection{CHANNEL CONDITION ASSUMPTIONS}\label{subsec:discussion_channels}
\ac{venena}'s security guarantee relies on the assumption that \emph{Eve} experiences statistically inferior channel conditions compared to \emph{Bob}. While this assumption does not hold universally, it reflects realistic deployment scenarios in modern wireless networks~\cite{CHZ+2025physical}. Fixed infrastructure transmitters (e.g., base stations) typically operate within secrecy-protected zones that prevent eavesdroppers from maintaining close proximity, either through physical security measures or detection mechanisms~\cite{ZGA+2011throughput,CCL+2014enhanced}. For mobile receivers, anonymity provisions in contemporary wireless protocols make precise localization challenging for adversaries, hindering their ability to consistently position themselves advantageously~\cite{BWB+2019modeling}. Furthermore, directional beamforming towards legitimate receivers naturally degrades eavesdropper channel quality when they cannot collocate with intended recipients. These factors collectively justify the channel asymmetry assumption for many practical scenarios including cellular downlink, wireless backhaul, and device-to-device communications. Nevertheless, scenarios where adversaries achieve sustained proximity to either transmitter or receiver (e.g., compromised infrastructure, targeted surveillance with dedicated resources) represent limitations of the current framework and motivate future research on adaptive security mechanisms that degrade gracefully under adverse channel conditions.
}

\revthree{
    \subsection{ADVANCED EAVESROPPERS}\label{subsec:advanced_eve}
    In this work, we have considered an informed \emph{Eve} with full knowledge of the system design, and verified \ac{venena}'s secrecy thereagainst. However, even more sophisticated eavesdropping strategies could be envisioned. For instance, \emph{Eve} can adapt its receiption and decryption strategy dynamically according to real-time channel conditions, which we have preliminarily explored in a generic and theoretical context of semantic \ac{pld} in~\cite{HZS+2025semantic}, and recently discussed in more depth in~\cite{CHZ+2025physical3}. Besides, multiple coordinated eavesdroppers may share their received signals and therewith overwhelm \emph{Bob} regarding the \ac{sinr}, in order to violate the secrecy. These advanced eavesdropping strategies pose new challenges to \ac{venena}, motivating future research on more resilient encryption strategies, which are out of the scope of this article.
}

\section{CONCLUSION AND OUTLOOKS}\label{sec:conclusion}
In this work, we have proposed \ac{venena}, a novel framework that combines physical layer deception with visual encryption and image poisoning for secure wireless transmission of semantics. By integrating vision transformer-based poisoning networks with power-domain multiplexing, we achieve effective deception of eavesdroppers while maintaining reliable communication with legitimate receivers.

{Experimental validation demonstrates} that \ac{venena} {maintains} above 93\% message perception accuracy for legitimate receivers while reducing eavesdroppers' {success rate by more than 40\%--even when adversaries possess full system knowledge. This quantifiable benchmark confirms the framework's active defense capability and practical robustness for 6G semantic communication scenarios.}

The sensitivity analysis {further identifies} optimal system configurations, particularly regarding the power mixing ratio {, and reveals \ac{venena}'s adaptability to diverse channel conditions. The discussion on efficiency, scalability, and channel assumptions provides guidance for future extensions, including multi-modal semantic encryption, adaptive mixing, and cross-layer deception mechanisms.}

With new possibilities in wireless physical layer security opened in this work, variate potential future topics are suggested towards future study. These are including but not limited to:
\begin{enumerate}
    \item the extension of the proposed framework to more complex visual content, such as high-resolution images that contain multiple semantic tags in each;
    \item the adoption to flexible semantic encoding that uses other forms of semantic information carriers, such as text or audio; and
    \item the development of dynamic adaptation strategies for challenging channel conditions.
\end{enumerate}

% \bibliographystyle{IEEEtran}
% \bibliography{references}

\end{document}

%% file: glossary.tex
\makeglossaries
\newacronym{5g}{5G}{Fifth Generation}
\newacronym{6g}{6G}{Sixth Generation}
\newacronym{ai}{AI}{artificial intelligence}
\newacronym{awgn}{AWGN}{additive white Gaussian noise}
\newacronym{bcd}{BCD}{block coordinate descent}
\newacronym{bpsk}{BPSK}{binary phase shift keying}
\newacronym{csi}{CSI}{channel state information}
\newacronym{csit}{CSIT}{channel state information at transmitter}
\newacronym{fbl}{FBL}{finite blocklength}
\newacronym{gan}{GAN}{generative adversarial network}
\newacronym{gma}{GMA}{gradient matching attack}
\newacronym{mimo}{MIMO}{multi-input multi-output}
\newacronym{nlos}{NLOS}{non-line-of-sight}
\newacronym{noma}{NOMA}{non-orthogonal multi-access}
\newacronym{nom}{NOM}{non-orthogonal multiplexing}
\newacronym{nve}{NVE}{Naive Visual Encryption}
\newacronym{ofdm}{OFDM}{orthogonal frequency-division multiplexing}
\newacronym{per}{PER}{packet error rate}
\newacronym{phy}{PHY}{physical}
\newacronym{pls}{PLS}{physical layer security}
\newacronym{pld}{PLD}{physical layer deception}
\newacronym{prb}{PRB}{physical resource block}
\newacronym{roc}{ROC}{receiver operating characteristic}
\newacronym{sic}{SIC}{successive interference cancellation}
\newacronym{sinr}{SINR}{signal-to-interference-and-noise ratio}
\newacronym{snr}{SNR}{signal-to-noise ratio}
\newacronym{venena}{VENENA}{Visual ENcryption for Eavesdropping NegAtion}
\newacronym{vit}{ViT}{vision transformer}